\documentclass[a4paper,11pt]{article}
\usepackage{bm}
\begin{document}
\title{Nonzero $\theta_{13}$ and CP Violation from Broken $\mu-\tau$ Symmetry}
\author{\bf{Asan Damanik}\footnote{E-mal: d.asan@lycos.com}\\Faculty of Science and Technology\\Sanata Dharma University\\Kampus III USD Paingan Maguwoharjo Sleman Yogyakarta\\Indonesia}
\date{}
\maketitle
\begin{abstract}
Nonzero and relatively large of $\theta_{13}$ mixing angle has some phenomenological consequences on neutrino physics beyond the standard model. One of the consequences if the mixing angle $\theta_{13}\neq 0$ is the possibility of the CP violation on the neutrino sector. In order to obtain nonzero $\theta_{13}$ mixing angle, we break the neutrino mass matrix that obey $\mu-\tau$ symmetry by introducing a complex parameter and determine the Jarlskog invariant as a measure of CP violation existence.  By using the experimental data as input, we determine the Dirac phase $\delta$ as function of mixing angle $\theta_{13}$.

\end{abstract}

\section{Introduction}
Recently, there are two unsolved major problems related to neutrino physics i.e. underlying symmetry of neutrino mass matrix and neutrino mixing matrix when confronted to the experimental results.  One of the interesting and popular underlying symmetry for neutrino mass matrix is the $\mu-\tau$ symmetry.  The $\mu-\tau$ symmetry reduces the number of parameters in symmetric neutrino mass matrix from 6 parameters to 4 parameters.  The neutrino mass matrix with $\mu-\tau$ symmetry can also obtained from three well-known neutrino mixing matrices i.e. tribimaximal (TBM), bimaximal (BM), dan democratic (DC).  But, the three well-known of neutrino mixing matrices predict mixing angle $\theta_{13}=0$ and hence Dirac phase $\delta$ can be arbitrary.  The standard parametrization of neutrino mixing matrix ($V$) is given by:
\begin{equation}
V=\bordermatrix{& & &\cr
&c_{12}c_{13} &s_{12}c_{13} &s_{13}e^{-i\delta}\cr
&-s_{12}c_{23}-c_{12}s_{23}s_{13}e^{i\delta} &c_{12}c_{23}-s_{12} s_{23}s_{13}e^{i\delta}&s_{23}c_{13}\cr
&s_{12}s_{23}-c_{12}c_{23}s_{13}e^{i\delta} &-c_{12}s_{23}-s_{12}c_{23}s_{13}e^{i\delta} &c_{23}c_{13}}
 \label{Vv}
\end{equation}
where $c_{ij}$ is the $\cos\theta_{ij}$, $s_{ij}$ is the $\sin\theta_{ij}$, and $\theta_{ij}$ are the mixing angles. In the basis where the charged lepton mass matrix is diagonal, the neutrino mass matrix $M_{\nu}$ can be diagonalized by mixing matrix $V$ as follow:
\begin{eqnarray}
M_{\nu}=VMV^{T},\label{m}
\end{eqnarray}
where the diagonal neutrino mass matrix $M$ is given by:
\begin{eqnarray}
M=\bordermatrix{& & &\cr
&m_{1} &0 &0\cr
&0 &m_{2} &0\cr
&0 &0 &m_{3}}. \label{Md}
\end{eqnarray}

If we put the mixing angle $\theta_{13}=0$ and consequently $s_{13}=0$ and $c_{13}=1$ in Eq. (\ref{Vv}), then the neutrino mass matrix $M_{\nu}$ in Eq. (\ref{m}) read:
\begin{eqnarray}
M_{\nu}=\bordermatrix{& & &\cr
&P &Q &Q\cr
&Q &R &S\cr
&Q &S &R},\label{M}
\end{eqnarray}
where:
\begin{eqnarray}
P=m_{1}c_{12}^{2}+m_{2}s_{12}^{2},\label{P1}\\
Q=(m_{2}-m_{1})c_{12}s_{12}c_{23},\label{Q1}\\
R=(m_{1}s_{12}^{2}+m_{2}c_{12}^{2})c_{23}^{2}+m_{3}s_{23}^{2},\label{R1}\\
S=(-m_{1}s_{12}^{2}-m_{2}c_{12}^{2}+m_{3})s_{23}c_{23}.\label{S1}
\end{eqnarray}

It is apparent from neutrino mass matrix ($M_{\nu}$) in Eq. (\ref{M}) that the neutrino mass matrix deduced from neutrino mixing matrix with assumption $\theta_{13}=0$ is $\mu-\tau$ symmetry.  But, as dictated from the experimental results, the mixing angle $\theta_{13}\neq 0$ and relatively large \cite{Double, Minos, T2K, Daya, RENO} which imlply that the asumption $\theta_{13}=0$ in formulating the neutrino mixing matrix must be rule out and hence the exact $\mu-\tau$ symmetry as the underlying symmetry of neutrino mass matrix is no longer adequate to accommodate the recent experimental results.

In order to accommodate the nonzero $\theta_{13}$ and Jarlskog rephasing invariant $J_{\rm CP}$ as a measure of CP violation in neutrino sector, in this paper we break the $\mu-\tau$ symmetry sofly by introducing one complex parameter to break the neutrino mass matrix with $\mu-\tau$ symmetry softly.  The paper is organized as follow: in section 2 we break the neutrino mass matrix with $\mu-\tau$ symmetry by introducing one complex parameter into neutrino mass matrix and in section 3 we determine the mixing angle $\theta_{13}$ and $J_{\rm CP}$ by using the experimental results as input.  Finally, section 4 is devoted to conclusions.

\section{Broken $\mu-\tau$ symmetry}
Concerning the neutrino mass matrix that obey the $\mu-\tau$  symmetry and mixing angle $\theta_{13}$, Mohapatra \cite{Mohapatra} stated explicitly that neutrino mass matrix that obey $\mu-\tau$ symmetry to be the reason for maximal $\mu-\tau$ mixing and one gets $\theta_{13}=0$, conversely if $\theta_{13}\neq 0$ can provide the $\mu-\tau$ symmetry beraking manifests in the case of normal hierarchy.  Aizawa and Yasue \cite{Aizawa} analysis complex neutrino mass texture and the $\mu-\tau$ symmetry which can yield small $\theta_{13}$ as a $\mu-\tau$ breaking effect.   The $\mu-\tau$ symmetry breaking effect in relation with the small $\theta_{13}$ also discussed in \cite{Fuki}.  Analysis of the correlation between CP violation and the $\mu-\tau$ symmetry breaking can be read in \cite{Mohapatra2, Baba, He1, Damanik}.

Now, we are in position to study the effect of neutrino mass matrix that obey the $\mu-\tau$ symmetry breaking in relation to the nonzero $\theta_{13}$ and Jarlskog rephasing invariant $J_{\rm CP}$ by breaking the neutrino mass matrix in Eq. (\ref{M}).  We break the neutrino mass matrix in Eq. (\ref{M}) by introducing a complex paramater $ix$ with the constraint that the trace of the broken neutrino mass matrix is remain constant or equal to the trace of the unbroken one.  This scenario of breaking has been applied by Damanik \cite{Damanik1} to break the neutrino mass matrix invariant under a cyclic permutation.  In this breaking scenario, the broken neutrino mass matrix reads:
\begin{equation}
M_{\nu}=\bordermatrix{& & &\cr
&P &Q &Q\cr
&Q &R-ix &S\cr
&Q &S &R+ix}.\label{M1}
\end{equation}

As stated previously that the CP violation can be determined from the Jarlskog rephasing invariant $J_{\rm CP}$.  Alternatively, Jarlskog rephasing invariant $J_{\rm CP}$ can be determined using the relation \cite{Branco}:
\begin{equation}
J_{\rm CP}=-\frac{{\rm Im}\left[(M_{\nu}^{'})_{e\mu}(M_{\nu}^{'})_{\mu\tau}(M_{\nu}^{'})_{\tau e}\right]}{\Delta m_{21}^{2}\Delta m_{32}^{2}\Delta m_{31}^{2}}, \label{JCP}
\end{equation}
where $(M_{\nu}^{'})_{ij}=(M_{\nu}M_{\nu}^{\dagger})_{ij}$ with $ i,j=e,\nu,\tau$, and $\Delta m_{kl}^{2}=m_{k}^{2}-m_{l}^{2}$ with $k=2,3,~ {\rm and}~ l=1,2$.  From Eq. (\ref{M1}), we have:
\begin{eqnarray}
M_{\nu}^{'}=\bordermatrix{& & &\cr
&P^{2}+2Q^{2} &Q(P+S+R+ix) &Q(P+R+S-ix)\cr
&Q(P+S+R-ix) &Q^{2}+R^{2}+S^{2}+x^{2} &Q^{2}+2S(R-ix)\cr
&Q(P+S+R+ix) &Q^{2}+2S(R+ix) &Q^{2}+R^{2}+S^{2}+x^{2}}. \label{M2}
\end{eqnarray}

\section{Nonzero $\theta_{13}$ and Jarlskog rephasing invariant}
From Eqs. (\ref{JCP}) and (\ref{M2}) we have the Jarlskog rephasing invariant as follow:
\begin{equation}
J_{\rm CP}=\frac{2Q^{2}\left[S(S+P)^{2}-Q^{2}(R+S+P)-R^{2}S\right]x-(2Q^{2}S)x^{3}}{\Delta m_{21}^{2}\Delta m_{32}^{2}\Delta m_{31}^{2}}. \label{JCP1}
\end{equation}
If we insert Eqs. (\ref{P1})-(\ref{S1}) into Eq. (\ref{JCP1}), then we have the $J_{\rm CP}$ as follow:
\begin{eqnarray}
J_{\rm CP}=\frac{2(m_{2}-m_{1})^{2}c_{12}^{2}s_{12}^{2}c_{23}^{2}}{\Delta m_{21}^{2}\Delta m_{32}^{2}\Delta m_{31}^{2}}[[(-m_{1}s_{12}^{2}-m_{2}c_{12}^{2}+m_{3})s_{23}c_{23}\nonumber\\
 \times((-m_{1}s_{12}^{2}-m_{2}c_{12}^{2}+m_{3})s_{23}c_{23}+m_{1}c_{12}^{2}+m_{2}s_{12}^{2})^{2} \nonumber\\
-(m_{2}-m_{1})^{2}c_{12}^{2}s_{12}^{2}c_{23}^{2}((m_{1}s_{12}^{2}+m_{2}c_{12}^{2})c_{23}^{2}+m_{2}s_{23}^{2}\nonumber\\
-(m_{1}s_{12}^{2}+m_{2}c_{12}^{2}-m_{3})s_{23}c_{23}+m_{1}c_{12}^{2}+m_{2}s_{12}^{2})\label{JCP2}\\ -((m_{1}s_{12}^{2}+m_{2}c_{12}^{2})c_{23}^{2}+m_{2}s_{23}^{2})^{2}(-m_{1}s_{12}^{2}-m_{2}c_{12}^{2}\nonumber\\
+m_{3})s_{23}c_{23}]x-c_{23}s_{23}[-m_{1}s_{12}^{2}-m_{2}c_{12}^{2}+m_{3}]x^{3}].\nonumber
\end{eqnarray}
From Eq. (\ref{JCP2}), one can see that the exact $\mu-\tau$ symmetry ($x=0$) lead to $J_{\rm CP}=0$ as claimed by many authors, and if we still want to have $J_{\rm CP}\neq 0$ from the $\mu-\tau$ symmetry, then we must break it softly.

From Eq. (\ref{JCP2}) one can see that in this breaking scenario, the Jarlskog rephasing invariant ($J_{\rm CP}$) in neutrino sector can be proceed if we break the neutrino mass matrix of $\mu-\tau$ symmetry by introducing complex parameter $ix$ parameter.  It is also apparent from Eq. (\ref{JCP2}) that $m_{1}\neq m_{2}$ as a constraint to the existence of the $J_{\rm CP}\neq 0$.

In order to get the value of Jarlskog rephasing invariant $J_{\rm CP}$ of Eq. (\ref{JCP2}), we use the experimental values of neutrino oscillation as input.  The current combined world data for neutrino squared-mass difference are given by \cite{Gonzales-Carcia, Fogli}:
\begin{eqnarray}
\Delta m_{21}^{2}=7.59\pm0.20 (_{-0.69}^{+0.61}) \times 10^{-5}~\rm{eV^{2}},\label{M21}\\
\Delta m_{32}^{2}=2.46\pm0.12(\pm0.37) \times 10^{-3}~\rm{eV^{2}},~\rm(for~ NH)\label{m32}\\
\Delta m_{32}^{2}=-2.36\pm0.11(\pm0.37) \times 10^{-3}~\rm{eV^{2}},~\rm(for~ IH)\label{m321}\\
\theta_{12}=34.5\pm1.0 (_{-2.8}^{3.2})^{o},~~\theta_{23}=42.8_{-2.9}^{+4.5}(_{-7.3}^{+10.7})^{o},~~\theta_{13}=5.1_{-3.3}^{+3.0}(\leq 12.0)^{o},
 \label{GD1}
\end{eqnarray}
at $1\sigma~(3\sigma)$ level.  We also put $m_{1}=0$ as a first approximation and within this scenario, for normal hierarchy (NH):
\begin{eqnarray}
m_{2}^{2}=\Delta m_{21}^{2},\\
m_{3}^{2}=\Delta m_{32}^{2}+\Delta m_{21}^{2}.\label{m3}
\end{eqnarray}

By inserting the valeus of Eqs. (\ref{M21})-(\ref{m3}), we have:
\begin{eqnarray}
J_{\rm CP}= 0.4644x-832.9790x^{3}.\label{JCP4}
\end{eqnarray}
From Eq. (\ref{JCP4}), we can determine the maximun value of $J_{\rm CP}$ by using the relation:
\begin{eqnarray}
\frac{dJ_{\rm CP}}{dx}=0,
\end{eqnarray}
which proceed $x=0.0167$.  By substituting the value of $x=0.0167$ into Eq. (\ref{JCP4}), we have the maximum value of Jarlskog rephasing invariant:
\begin{eqnarray}
J_{\rm CP}\approx 0.004.\label{JCPm}
\end{eqnarray}

It is also possible to determine the Jarlskog rephasing invariant $J_{\rm CP}$ from the neutrino mixing matrix by using the relation: 
\begin{eqnarray}
J_{\rm CP}={\rm Im} (V_{11}^{*}V_{23}^{*}V_{13}V_{21}).\label{ADD}
\end{eqnarray}
From neutrino mixing matrix of Eq. (\ref{Vv}), the Jarlskog rephasing invariant $J_{\rm CP}$ reads:
\begin{eqnarray}
&J_{\rm CP}=c_{12}s_{12}c_{23}s_{23}c_{13}^{2}s_{13}\sin{\delta}\nonumber\\
&~~~~~~~~~~~~=c_{12}s_{12}c_{23}s_{23}\left(s_{13}-s_{13}^{3}\right)\sin{\delta}.\label{AD1}
\end{eqnarray}
If we put the experimental values of mixing angles $\theta_{12}$ and $\theta_{23}$ of Eq. (\ref{GD1}) into Eq. (\ref{AD1}), then we have:
\begin{eqnarray}
J_{\rm CP}=0.2327\left(s_{13}-s_{13}^{3}\right)\sin{\delta}.\label{ADD2}
\end{eqnarray}

Since the value of $s_{13}^{3}<<s_{13}$, the Eq. (\ref{ADD2}) can be approximated as follow:
\begin{eqnarray}
J_{\rm CP}\approx 0.2327s_{13}\sin{\delta}.\label{AD3}
\end{eqnarray}
If we insert the value of Jarlskog rephasing invariant of Eq. (\ref{JCPm}) into Eq.  (\ref{AD3}), then we have the Dirac phase $\delta$ as follow:
\begin{eqnarray}
\delta\approx\arcsin\left(\frac{0.0172}{\sin\theta_{13}}\right).
\end{eqnarray}
By inserting the value of mixing angle $\theta_{13}$ as shown in Eq. (\ref{GD1}), we have:
\begin{eqnarray}
\delta\approx 11.2^{\rm o}
\end{eqnarray}
\section{Conclusions}
We have studied sistematically the effect of breaking on neutrino mass matrix that obey $\mu-\tau$ symmetry by introducing a complex parameter $ix$ with the requirement that the trace of the broken $\mu-\tau$ symmetry is remain constant.  By using the experimental data as input, we can obtain the Jarlskog rephasing invariant $J_{\rm CP}\neq 0$ that indicate the existence of CP violation in neutrino sector and hence the Dirac phase $\delta$ which also depend on the mixing angle $\theta_{13}$ for neutrino in normal hierarchy for the case: $m_{1}=0$.

\end{document}